\documentclass[journal,article,submit,pdftex,moreauthors]{Definitions/mdpi} 
\firstpage{1} 
\pubvolume{1}
\issuenum{1}
\articlenumber{0}
\pubyear{2022}
\copyrightyear{2022}
\datereceived{} 
\dateaccepted{} 
\datepublished{} 
\hreflink{https://doi.org/} 

\Title{Measurement of the central Galactic black hole by extremely large mass-ratio inspirals}

\TitleCitation{Title}

\Author{Shu-Cheng Yang $^{1,\dagger}$*\orcidA{}, Huijiao Luo $^{1,}$, Yuan-Hao Zhang $^{2,3}$ and Chen Zhang $^{1,}$}

\AuthorNames{Firstname Lastname, Firstname Lastname and Firstname Lastname}

\AuthorCitation{Lastname, F.; Lastname, F.; Lastname, F.}

\address{%
$^{1}$ \quad Shanghai Astronomical Observatory, Chinese Academy of Sciences, Shanghai 200030, China\\
$^{2}$ \quad School of Fundamental Physics and Mathematical Sciences, Hangzhou Institute for Advanced Study, UCAS, Hangzhou 310024, China\\

$^{3}$ \quad University of Chinese Academy of Sciences (UCAS), Beijing 100049, China}

\usepackage{color}

\corres{Correspondence: ysc@shao.ac.cn}

\abstract{In the Galaxy, extremely large mass-ratio inspirals(X-MRIs) composed of brown dwarfs
and the massive black hole at the Galactic Center are expected to be promising gravitational wave sources for space-borne detectors. In this work, we simulate the gravitational wave signals from twenty X-MRI systems by an axisymmetric Konoplya-Rezzolla-Zhidenko metric with varied parameters. We find that the mass, spin, and deviation parameters of the Kerr black hole could be determined accurately ( $\sim 10^{-5} - 10^{-6}$ ) with only one X-MRI event with a high signal-to-noise ratio. The measurement of the above parameters could be improved with more X-MRI observations.}


\begin{document}




\section{Introduction}


The first observations of gravitational waves(GWs) from binary black hole mergers and binary neutron star inspirals ushered in a new era of GW physics and astronomy\cite{ligo2016observation, ligo2017gw170817}. Since then, the ground-based detectors have detected 90 GW events\cite{ligo2019gwtc1, ligo2021gwtc2, ligo2021gwtc3}. The detectable frequency band of current ground-based GW detectors such as Advanced LIGO\cite{ligo2015advanced}, Advanced Virgo \cite{acernese2014advanced}, and KAGRA \cite{kagra2019kagra} ranges from 10 to 10,000 Hz, which makes ground-based GW detectors unable to detect any GWs with frequencies less than 10 Hz, while abundant sources are emitting GWs in the low-frequency band\cite{amaro2007intermediate}. The space-borne GW detectors such as LISA \cite{amaro2017laser}, Taiji \cite{hu2017taiji}, and TianQin \cite{luo2016tianqin}, which will be launched in the 2030s, will open GW windows from 0.1 mHz to 1 Hz, and are expected to probe the nature of astrophysics, cosmology, and fundamental physics.

One of the most essential and promising GW sources for space-borne GW detectors is the extreme-mass ratio inspiral (EMRI), which is formed when a massive black hole (MBH) captures a small compact object.\cite{amaro2007intermediate, gair2010lisa}. The word "inspiral" here means the inspiralling process that the relatively lighter object gradually spirals in toward the MBH due to the emission of GWs. The small object should be compact to keep it from being tidally disrupted by the MBH so that it is unlikely to be a main-sequence star. The possible candidate could be a stellar-mass black hole(BH),  neutron star,  white dwarf, or other compact objects. The designed space-borne detectors will be sensitive to EMRIs that contain MBHs with the mass $10^{4}-10^{7}M_{\odot}$ and small compact objects with stellar mass, and the fiducial mass ratio will be $10^{3} - 10^{6}$\cite{chua2017augmented}. 

Moreover, a special kind of EMRI, extremely large mass-ratio inspirals (X-MRIs) with a mass ratio of $q\sim10^{8}$ also are potential sources for space-borne GW detectors\cite{gourgoulhon2019gravitational, amaro2019extremely}. The X-MRI system is formed when an MBH captures a brown dwarf (BD) with mass $\sim10^{-2}~ M_{\odot}$. Brown dwarfs are substellar objects with insufficient mass to sustain nuclear fusion and become main-sequence stars\cite{burrows1993science}. Brown dwarfs are denser than main-sequence stars, and their Roche limit is closer to the horizon of MBH \cite{freitag2002gravitational,gourgoulhon2019gravitational}. Therefore,  brown dwarfs could survive very close to the MBH. 

The mass of BD is relatively tiny, so space-borne GW detectors like LISA could only observe X-MRIs nearby, especially X-MRIs at the Galactic Center(GC)\cite{gourgoulhon2019gravitational, amaro2019extremely}. The MBH of these X-MRIs, Sgr A*, is 8 kpc from the solar system, and its mass is about $4\times 10^6~M_{\odot}$ \cite{eckart1996observations,ghez1998high,ghez2008measuring, genzel2010galactic}. A typical X-MRI at the GC covers $\sim10^{8}$ cycles,  which last millions of years in the LISA band\cite{amaro2019extremely}. Such X-MRI could have a relatively high SNR (more than 1000), and dozens of X-MRIs might be observed during the LISA mission period\cite{amaro2019extremely}. Therefore, the X-MRIs at the GC offers a natural laboratory for studying the properties of BH and testing theories of gravity. 

In this paper, we simulate the GW signals of X-MRIs at GC to show how and to what extent the fine structure of Sgr A* could be figured. In general relativity(GR), according to the no-hair theorem, BHs are characterized by their masses, spins, and electric charges, and the Kerr metric is believed to be the metric that describes the space-time of BH. However, alternative theories of gravity predict hairy black holes \cite{afrin2021parameter} and other  metrics that describe the space-time of BH\cite{konoplya2016general}. The parameterized metrics are proposed to describe the space-time of non-Kerr black holes. In this paper, to describe the space-time of X-MRIs at GC, we use a model-independent parameterization metric, Konoplya-Rezzolla-Zhidenko metric(KRZ metric)\cite{konoplya2016general}, which can describe metrics that is generic stationery and axisymmetric.

This paper is organized as follows, in section \ref{KRZ}, we review the KRZ parametrization. In section \ref{Kludge}, we introduce the "kludge" waveforms used in our work and simulate the GW signals emitted by X-MRIs at GC. In section \ref{data}, based on the simulated GW signals, we apply the Fisher matrix to these GWs and present the accuracy of parameter estimation of Sgr A* for future space-borne GW detectors. The conclusion and outlook are given in section\ref{sec:conclusion}. Throughout this letter, we use natural units $(G=c=1)$, greek letters $(\mu,\nu,\sigma,...)$ stand for space-time indices, and Einstein summation is assumed.




\section{KRZ prametrized metric}
\label{KRZ}
GR is the most accurate and concise theory of gravity by far\cite{abbott2021tests}.
While in practice, there are quite a few other theories of gravity, whose predictions resemble general relativity's, to be tested. In the framework of GR, the Schwarzschild or Kerr metric describes the space-time of uncharged BH. However, in modified and alternative theories of gravity, there are other possible solutions for the description of the space-time of BHs\cite{hu2022observational, cao2022integrability, zhang2022equivalence, yang2022chaos,  zhang2021charged, yi2020dynamics}. The predictions of different theories of gravity are different, so a universal and reasonable theory about the GWs of X-MRI should be model-independent.

In order to deal with numerous metrics of non-Kerr black holes, one may use the parameterized metric to describe the space-time of non-Kerr black holes. There are several model-independent frameworks, one of which parametrizes the most generic black hole geometry through a finite number of adjustable quantities and is known as Johannsen-Psaltis parametrization (J-P metric)~\cite{johannsen2011metric}. The J-P metric expresses deviations from general relativity in terms of a Taylor expansion in powers of $M/r$, where $M$ is the mass of BH  and $r$ is the radial coordinate.The J-P parametrization is widely adopted, but it is not a robust and generic parametrization for rotating black holes~\cite{konoplya2016general,ni2016testing}. Notably, the parametric axisymmetric J-P metric obtained from the Janis-Newman algorithm~\cite{drake2000uniqueness} does not cover all deviations from Kerr space-time.

Another model-independent parameterization metric ~\cite{konoplya2016general, jiang2015using}, KRZ metric, is based on a double expansion in both the polar and radial directions of a generic stationary and axisymmetric metric.The KRZ metric is effective in reproducing the space-time of three commonly used rotating black holes (Kerr, rotating dilation\cite{horne1992rotating}, and Einstein-dilaton-Gauss-Bonnet black holes\cite{cardoso2014generic}) with finite parameters (see Ref.\cite{konoplya2016general} for more details). According to KRZ parameterization,  the space-time of any axisymmetric black hole with total mass $M$ and rotation parameter $a$ could be expressed in the following form\cite{konoplya2016general}:
\begin{eqnarray}
\label{E1}
    ds^{2}&=&-\frac{N^{2}-W^{2}\sin^{2}\theta}{K^{2}}dt^{2}-2Wr\sin^{2}\theta dtd\phi\nonumber\\ && +K^{2}r^{2}\sin^{2}\theta d\phi^{2}+S\left(\frac{B^{2}}{N^{2}}dr^{2}+r^{2}d\theta^{2}\right),
\end{eqnarray}
where \cite{younsi2016new}
\begin{eqnarray}
\label{E2}
    S&=&\frac{\Sigma}{r^{2}}=1+\frac{a^{2}}{r^{2}}\cos^{2}\theta \ , 
\\
    \Sigma&=&r^{2}+a^{2}\cos^{2}\theta , 
\end{eqnarray}
$N, B, W$,and $K$ are the functions of the radial and polar coordinates (expanded in term $\cos\theta$), 
\begin{eqnarray}
\label{E3}
    W&=&\sum_{i=0}^\infty\frac{W_{i}(r)(\cos\theta)^{i}}{S},\\
    B&=&1+\sum_{i=0}^\infty B_{i}(r)(\cos\theta)^{i},\\
    N^{2}&=&\left(1-\frac{r_{0}}{r}\right)A_{0}(r)+\sum_{i=1}^\infty A_{i}(r)(\cos\theta)^{i},\\
    K^{2}&=&1+\frac{aW}{r}+\frac{a^{2}}{r^{2}}+\sum_{i=1}^\infty\frac{K_{i}(r)(\cos\theta)^{i}}{S}, \label{E3k}
\end{eqnarray}
with 
\begin{eqnarray}
    &&B_{i}=b_{i0}\frac{r_{0}}{r}+\tilde{B_{i}}\frac{r_{0}^{2}}{r^{2}},\\
    &&\tilde{B_{i}}\equiv\frac{b_{i 1}}{1+\frac{x b_{i2}}{1+\frac{x b_{i3}}{1+...}}},\\
    &&W_{i}=b_{i0}\frac{r_{0}^{2}}{r^{2}}+\tilde{B_{i}}\frac{r_{0}^{3}}{r^{3}},\\
    &&\tilde{W_{i}}\equiv\frac{\omega_{i 1}}{1+\frac{x\omega_{i2}}{1+\frac{x\omega_{i3}}{1+...}}},\\
    &&K_{i>0}(r)=k_{i0}\frac{r_{0}^{2}}{r^{2}}+\tilde{K_{i}}\frac{r_{0}^{3}}{r^{3}},\\
    &&\tilde{K_{i}}\equiv\frac{k_{i 1}}{1+\frac{x k_{i2}}{1+\frac{x k_{i3}}{1+...}}},
\\
    &&A_{0}(r)=1-\epsilon_{0}\frac{r_{0}}{r}+(a_{00}-\epsilon)\frac{r_{0}^{2}}{r^{2}}+\frac{a^{2}}{r^{2}}+\tilde{A_{0}}\frac{r_{0}^{3}}{r^{3}},\\
    &&A_{i>0}=K_{i}(r)+\epsilon_{i}\frac{r_{0}^{2}}{r^{2}}+a_{i0}\frac{r_{0}^{3}}{r^{3}}+\tilde{A_{i}}\frac{r_{0}^{4}}{r^{4}},\ \ \\
    &&\tilde{A_{i}}\equiv\frac{a_{i 1}}{1+\frac{x a_{i2}}{1+\frac{x a_{i3}}{1+...}}}, 
\label{Adefs}
\end{eqnarray}
where $x=1-{r_{0}}/{r}$, and $r_{0}$ is the radius of the black hole horizon in the equatorial plane. The metric (\ref{E1}) is characterized by the order of expansion in radial and polar directions. The parameters $a_{i j},b_{i j},\omega_{i j},k_{i j}$ (here $i=0,1,2,3..., j=1,2,3...$) are effectively independent. This is because one of these functions, $A_{i}(x), B_{i}(x), W_{i}(x)$ and $K_{i}(x)$, is fixed by coordinate choice\cite{younsi2016new}. 

In the following, we present the parameterized metric with first-order radial expansion and second-order polar direction, which describe the space-time of a deformed Kerr black hole\cite{younsi2016new, ni2016testing}:

\begin{eqnarray}
    B&=&1+\frac{\delta_{4}r_{0}^{2}}{r^{2}}+\frac{\delta_{5}r_{0}^{2}}{r^{2}}\cos^{2}\theta,\\
    W&=&\frac{1}{\Sigma}\left[\frac{\omega_{00}r_{0}^{2}}{r^{2}}+\frac{\delta_{2}r_{0}^{3}}{r^{3}}+\frac{\delta_{3}r_{0}^{3}}{r^{3}}\cos^{2}\theta\right],\\
    K^{2}&=&1+\frac{aW}{r}+\frac{1}{\Sigma}\left(\frac{k_{00}r_{0}^{2}}{r^{2}}+\frac{k_{21}r_{0}^{3}}{r^{3}}\cos^{2}\theta\right),\\
    N^{2}&=&\left(1-\frac{r_{0}}{r}\right)\left[1-\frac{\epsilon_{0}r_{0}}{r}+(k_{00}-\epsilon_{0})\frac{r_{0}^{2}}{r^{2}}+\frac{\delta_{1}r_{0}^{3}}{r^{3}}\right]\nonumber \\ && +\left[(k_{21}+a_{20})\frac{r_{0}^{3}}{r^{3}}+\frac{a_{21}r_{0}^{4}}{r^{4}}\right]\cos^{2}\theta,
\end{eqnarray}

The radius of the horizon and the Kerr parameter are 
\begin{equation}
\label{E4}
    r_{0}=M+\sqrt{M^{2}-a^{2}}, \ \ a=J/M,
\end{equation}
where $J$ is the total angular momentum. For simplicity, here $M$ has one unit, i.e. $ M = 1$. One can obtain related variables and parameters from dimensionless quantity by scale transformations\cite{zhou2022note, wang2021construction, wang2021construction2, wang2021construction3, wu2021construction4, sun2021applying}: $tM \rightarrow t$, $rM \rightarrow r$, etc. The coefficient $r_{0}$, $a_{20}$, $a_{21}$, $\epsilon_{0}$, $k_{00}$, $k_{21}$ and $\omega_{00}$ in the KRZ metric can be expressed as follows\cite{ni2016testing, xin2019gravitational}
\begin{eqnarray}
    r_{0}&=&1+\sqrt{1-\tilde{a}^{2}}, \\
     a_{20}&=&\frac{2\tilde{a}^{2}}{r_{0}^{3}},\\
    a_{21}&=&-\frac{\tilde{a}^{4}}{r_{0}^{4}}+\delta_{6},\\
    \epsilon_{0}&=&\frac{2-r_{0}}{r_{0}},\\
     \omega_{00}&=&\frac{2\tilde{a}}{r_{0}^{2}},\\
    k_{00}&=&\frac{\tilde{a}^{2}}{r_{0}^{2}},\\
    k_{21}&=&\tilde{a}^{4}/r_{0}^{4}-2\tilde{a}^{2}/r_{0}^{3}-\delta_{6},\\
    k_{22}&=&-\tilde{a}^{2}/r_{0}^{2}+\delta_{7},\\
    k_{23}&=&\tilde{a}^{2}/r_{0}^{2}+\delta_{8},
\end{eqnarray}
here $\tilde{a}=a/M$ stands for the spin parameter. The deformation parameters ${\delta_{j}}(j=1,2,...,8)$ represent the deviations from the Kerr metric. The physical meaning of these parameters could be summarized as follows: $\delta_{1}$ is related to deformation of $g_{tt}$; $\delta_{2},\delta_{3}$ are related to the rotational deformation of the metric; $\delta_{4},\delta_{5}$ are related to deformation of $g_{rr}$ and $\delta_{6}$ is related to the deformation of the event horizon (see Ref. \cite{konoplya2016general} for more details). The KRZ metric is an appropriate tool to measure the potential deviations from the Kerr metric. As a first order approximation, in this work we mainly consider $\delta_{1}$ and $\delta_{2}$.

\section{Waveform model for KRZ black holes}
\label{Kludge}

Several waveform models can simulate the signal of EMRI\cite{hughes2001evolution,barack2004lisa,drasco2006gravitational,babak2007kludge,chua2015improved,chua2017augmented}. Among these models, the kludge model can generate waveforms quickly and have a 95\% accuracy compared with the Teukolsky-based waveforms\cite{babak2007kludge}. The kludge waveforms may be essential in searching for EMRIs/X-MRIs for future space-borne GW detectors. We employ the kludge waveforms to simulate X-MRI waveforms\cite{xin2019gravitational}. Before presenting the results, we would like to review the structure and logic of the calculation. The calculation of waveforms can be summarized in the following steps:  

\begin{itemize}
    \item First, to consider the brown dwarf of the X-MRI as a point particle. 
    \item Second, to use the given metric to calculate the particle's trajectory by integrating the geodesic equations that contain the radiation flux. 
    \item Finally, to use the quadrupole expression to get the GWs emitted from the system of the X-MRI.
\end{itemize}

To get the trajectory of the particle, we start by calculating the geodesics using the following equations:
\begin{eqnarray}
\label{Eeins1}
    \dot{u}^{\mu}&=&-\Gamma_{\rho\sigma}^{\mu}u^{\rho}u^{\sigma}, \\
    \dot{x}^{\mu}&=&u^{\mu}, 
\end{eqnarray}
where $x^{\mu}$ is the coordinate of the particle, ${u}^{\mu}$ is the 4-velocity, which satisfies
\begin{equation}
|u| = g_{\mu\nu}u^{\mu}u^{\nu}=-1\ ,
\end{equation}
and $\Gamma_{\rho\sigma}^{\mu}$ are the Christoffel symbols. For stable bounded geodesics, the orbital eccentricity $e$ and semi-latus rectum $p$ can be defined by periastron $r_{p}$ and apastron $r_{a}$, and the inclination angle $\iota$ is defined in the Keplerian convention by:
\begin{align}
    e&=\frac{r_{a}-r_{p}}{r_{a}+r_{p}},
    &p&=\frac{2r_{a}r_{p}}{r_{a}+r_{p}},
    &\iota&=\frac{\pi}{2}-\theta_{\mathbf{min}}.
\end{align}
where $\theta_{\mathbf{min}}$ is the minimum of $\theta$ along the geodesic. The geodesic may be specified by the parameters $(r_{a},r_{p},\theta_{\mathbf{min}})$, which fully describe the range of motion in the radial and polar coordinates. In this paper, we define $(e, p, \iota)$ from $(r_a, r_p, \theta_{\mathbf{min}})$ by the numerically generated trajectory.

In the background of Kerr metric, the geodesic can be described by the orbital energy $E$, the $z$ component of the orbital angular momentum $L_{z}$, and the Carter constant $Q$\cite{xin2019gravitational}. $E$ and $L_{z}$ still exist in the KRZ background, and take the form

\begin{eqnarray}
\label{u}
    E&=&-u_{t}=-g_{tt}u^{t}-g_{t\phi}u^{\phi}\ ,
\\
    L_{z}&=&u_{\phi}=g_{t\phi}u^{t}+g_{\phi\phi}u^{\phi}\ .
\end{eqnarray}
Strictly speaking, unlike the Kerr metric, the Carter constant $Q$ does not exist in the KRZ metric. While when considering the situations that are close to the Kerr metric, we use an approximate "Carter constant" \cite{rudiger1981conserved, rudiger1983conserved}
\begin{align}
\label{Q}
Q=L_z^{2}\tan^{2}\iota,
\end{align}
While the orbital constants $(E,L_{z},Q)$ in the above geodesic setup do not vary with time, it is convenient to work with alternative parametrizations of $(E,L_{z},Q)$. The relationship between $(r_{a},r_{p},\theta_{\mathbf{min}})$ and $(E,L_{z},Q)$ is given by\cite{chua2017augmented}
\begin{eqnarray}
     &&P^{2}{|_{r=r_{a}, \theta=\pi/2}}-[r^{2}+(L_{z}^{2}-aE)^{2}+Q]\Delta{|_{r=r_{a}, \theta=\pi/2}}=0,\qquad\\
     &&P^{2}{|_{r=r_{p}, \theta=\pi/2}}-[r^{2}+(L_{z}^{2}-aE)^{2}+Q]\Delta{|_{r=r_{p}, \theta=\pi/2}}=0,\\
     &&Q=\cos^{2}\theta_{\rm min}\left[a^{2}(1-E^{2})+\left(\frac{L_{z}}{\sin\theta}\right)^2\right].
\end{eqnarray}

Because of the extreme mass ratio of X-MRI, the deviations from the geodesics due to radiation reaction should be small. While in this work, for accuracy, we consider the effect of radiation reaction, which is included by replacing the Eq.~(\ref{Eeins1}) with the following one 
\begin{equation}
    \frac{du^{\mu}}{d\tau}=-\Gamma_{\rho\sigma}^{\mu}u^{\rho}u^{\sigma}+\mathcal{F}^{\mu}
\end{equation}
where the radiation force $\mathcal{F}^{\mu}$ is connected with the adiabatic radiation fluxes $(\dot{E},\dot{L_{z}},\dot{Q})$ as
\begin{eqnarray}
\label{Flux}
&\dot{E}u^{t}=-g_{tt}\mathcal{F}^{t}-g_{t\phi}\mathcal{F}^{\phi},\\
&\dot{L_{z}}u^{t}=g_{t\phi}\mathcal{F}^{t}+g_{\phi\phi}\mathcal{F}^{\phi},\\
&\dot{Q}u^{t}=2g_{\theta\theta}^{2}u^{\theta}\mathcal{F}^{\theta}+2\cos^{2}\theta^{2}a^{2}E\dot{E}+2\cos^{2}\theta\frac{L_{z}\dot{L_{z}}}{\sin^{2}\theta},\\
&g_{\mu\nu}u^{\mu}\mathcal{F}^{\nu}=0.
\end{eqnarray}
Eq.~(\ref{Flux}) can be deduced by taking derivatives with respect to proper time in Eqs.~(\ref{u})-(\ref{Q}). Integrating the geodesic equations that contain the radiation flux is crucial for calculating the particle's trajectory. In this paper, due to the short integration time, we use the Runge-Kutta method. There are also several geometric numerical integration methods for integrating the equations of geodesics. Such as manifold correction schemes\cite{wang2016implementation, wang2018simulations, deng2020use}, extended phase space methods\cite{li2017modification, luo2017explicit, pan2021extended, liu2016higher}, explicit and implicit combined symplectic methods\cite{mei2013preference, mei2013dynamics, zhong2010global}, and explicit symplectic integrators\cite{zhou2022note, wang2021construction, wang2021construction2, wang2021construction3, wu2021construction4, sun2021applying}. For situations such as the long-term evolution of Hamiltonian systems\cite{deng2020use}, geometric numerical integration methods can be helpful. 

Finally, after generating the trajectory, we turn to the third step -- to calculate the gravitational waveforms. We start from transforming the Boyer-Lindquist coordinates $(t,r,\phi,\theta)$ into Cartesian coordinates $(t,x,y,z)$ using the relations: 
\begin{eqnarray}
    t&=&t,\\
    x&=&r\sin\theta\cos\phi,\\
    y&=&r\sin\theta\sin\phi,\\
    z&=&r\cos\theta.
\end{eqnarray}
Then we calculate the quadrupole expression (see Ref.~\cite{babak2007kludge})

\begin{eqnarray}
\label{Quadrupole}
&&    \bar{h}^{jk}(t,x)=\frac{2}{r}[\ddot{I}^{jk}(t^{'})]_{t^{'}=t-r},\\
&&    I^{jk}(t^{'})=\int x^{'j}x^{'k}T^{00}(t^{'},x^{'})d^{3}x^{'},
\end{eqnarray}
where $I^{jk}(t^{'})$ is the source's mass quadrupole moment, $T^{00}$ is component of the energy-momentum tensor $T^{\mu\nu}(t^{'},x^{'})$, and $\bar{h}^{\mu\nu}=h^{\mu\nu}-\frac{1}{2}\eta^{\mu\nu}\eta^{\rho\sigma}h_{\rho\sigma}$ is the trace-reversed metric perturbation.
Then we transform the waveform into the transverse-traceless gauge (see Ref.~\cite{babak2007kludge} for more details) 
\begin{equation}
h_{TT}^{jk}=\frac{1}{2}
\begin{pmatrix}
0&0&0\\
0&h^{\Theta\Theta}-h^{\Phi\Phi}&2h^{\Theta\Phi}\\
0&2h^{\Theta\Phi}&h^{\Phi\Phi}-h^{\Theta\Theta}
\end{pmatrix} \ , 
\end{equation}
with
\begin{eqnarray}
    h^{\Theta\Theta}&=&\cos^{2}\Theta\left[h^{xx}\cos^{2}\Phi+h^{xy}\sin2\Phi+ h^{yy}\sin^{2}\Phi\right]\nonumber\\ && +h^{zz}\sin^{2}\Theta-\sin2\Theta\left[h^{xz}\cos\Phi+h^{yz}\sin\Phi\right],\\
    h^{\Phi\Theta}&=&\cos\Theta\left[-\frac{1}{2}h^{xx}\sin2\Phi+h^{xy}\cos2\Phi+\frac{1}{2}h^{yy}\sin2\Phi\right]\nonumber \\ &&+\sin\Theta\left[h^{xz}\sin\Phi-h^{yz}\cos\Phi\right],\\
    h^{\Phi\Phi}&=&\left[h^{xx}\sin^{2}\Phi-h^{xy}\sin2\Phi+h^{yy}\cos^{2}\Phi\right].
\end{eqnarray}
Now we get the plus and cross components of the waveform observed at latitudinal angle $\Theta$ and azimuthal angle $\Phi$
\begin{eqnarray}
    h_{+}&=&h^{\Theta\Theta}-h^{\Phi\Phi}\nonumber \\
        &=&\cos^{2}\Theta\left[h^{xx}\cos^{2}\Phi+h^{xy}\sin2\Phi+ h^{yy}\sin^{2}\Phi\right]\nonumber \\&&+h^{zz}\sin^{2}\Theta-\sin2\Theta\left[h^{xz}\cos\Phi+h^{yz}\sin\Phi\right]\nonumber \\
        &&-\left[h^{xx}\sin^{2}\Phi-h^{xy}\sin2\Phi+h^{yy}\cos^{2}\Phi\right], \label{hplus}\\ 
    h_{\times}&=&2h^{\Theta\Phi}\nonumber \\
    &=&2\Big\{\cos\Theta\left[-\frac{1}{2}h^{xx}\sin2\Phi+h^{xy}\cos2\Phi+\frac{1}{2}h^{yy}\sin2\Phi\right]\nonumber \\ && +\sin\Theta\left[h^{xz}\sin\Phi-h^{yz}\cos\Phi\right]\Big\}.  \label{hcross}
\end{eqnarray}

The strength of the
signal in a detector could be characterized by the signal-to-noise ratio (SNR). The SNR of the signals can be defined as\cite{finn1992detection}
\begin{equation}
\label{SNR}
    \rho:=\sqrt{\left\langle h\vert h\right\rangle}, 
\end{equation}
 where $\left\langle \cdot\vert \cdot\right\rangle$ is the standard matched-filtering inner product between two data streams. The inner product between signal $a(t)$ and template $b(t)$ is

\begin{equation}
    \left\langle a\vert b\right\rangle=2\int_{0}^{\infty}\frac{\tilde{a}^{*}(f)\tilde{b}(f)+{\tilde{a}(f)\tilde{b}^{*}(f)}}{S_{n}(f)}df
\end{equation}
where $\tilde{a}(f)$ is the Fourier transform of the time series signal $a(t)$, $\tilde{a}^{*}(f)$ is the complex conjugate of $\tilde{a}(f)$ and $S_{n}(f)$ is the power spectral density of the GW detectors' noise. Throughout this paper, the power spectral density is taken to be the noise level of LISA.

In this work, to quantify the differences between GW signals and the templates, we use maximized fitting factor (overlap)
\begin{equation}
    \mathrm{FF}(a,b)=\frac{(a\vert b)}{\sqrt{(a\vert a)(b\vert b)}}.
\end{equation}
If we include the time shift $t_{s}$ and the phase shift $\phi_{s}$, the fitting factor reads
\begin{equation}
    \mathrm{ff}(t_{s},\phi_{s},a(t),b(t))=\frac{(a(t)\vert b(t+t_{s})e^{i\phi_{s}})}{\sqrt{(a\vert a)(b\vert b)}},
\end{equation}
the maximized fitting factor is defined as 
\begin{equation}
    \mathrm{FF}(a,b)=\underset{t_{s},\phi_{s}}{\mathbf{max}}\frac{(a(t)\vert b(t+t_{s})e^{i\phi_{s}})}{\sqrt{(a\vert a)(b\vert b)}}. 
\end{equation}

\section{Data analysis}
\label{data}
In this section, we first specify the main parameters values we used in this work. Then we use XSPEG, a software for generating GWs in the KRZ metric, provided by the authors of Ref.~\cite{xin2019gravitational} to calculate the gravitational waveforms and do some analysis. Finally, we employ the Fisher information matrix to evaluate the parameter estimation accuracy for LISA-like GW detectors.

For X-MRI  at the GC, the mass of the brown dwarfs ranges from $\sim 0.01~\rm M_{\odot}$ to $\sim 0.08~\rm M_{\odot}$ \cite{chabrier2000theory}. The parameter values for the MBH Sgr A* in this work are as follows:
\begin{itemize}
    \item the mass of Sgr A* $M_{\rm Sgr A*}=4\times10^{6}~\rm {M_{\odot}}$\cite{eckart1996observations,ghez1998high,ghez2008measuring}; 
    
    \item the dimensionless spin parameter $a=0.5$ \cite{shcherbakov2012sagittarius};
    \item the distance between Sgr A* and the solar system $R_{p}=8.3$ kpc \cite{eisenhauer2003geometric};
    \item the latitudinal angle $\Theta=-29^{\circ}$ and the azimuthal angle $\Phi=266.417^{\circ}$ \cite{menten1997position}. 
\end{itemize}

Based on the parameters above, we first simulate the GW signals of twenty X-MRIs at the GC (see Table \ref{tab:parameters} ). The mass ratio $q$ ranges from $5\times10^{7}$ to $4.0\times10^{8}$, the orbit eccentricity $e$ ranges from $0.1$ to $0.8$, the semi-latus rectum $p$ ranges from $10.6$ to $50.0$, the inclination angle $\iota$ ranges from $-2\pi/3$ to $\pi/3$, and the duration of above signals is one year. Then, we calculate the overlaps between above GW signals and many GW series with varying parameters. Finally, we use the Fisher information matrix to provide the uncertainties of parameter estimations.

\subsection{The overlaps between simulated GW signals of XMRIs and GW series with varying parameters}
\label{A}

\begin{table*}[!htbp]
    \centering
    \resizebox{\textwidth}{!}{
    \begin{tabular}{ccccccccccc}
    \toprule
        Signal&e&p&$\iota$&$M_{Object}$&SNR&$\Delta a/a$&$\Delta M/M$&$\Delta\delta_{1}$&$\Delta\delta_{2}$&$\Delta R_{p}/R_{p}$\\
        \hline
        $01$ &  $0.617$ & $10.600$ & $5\pi/6$ & $2.80\times10^{-2}$ & $1584.363$ &$2.85\times10^{-6}$    &$4.18\times10^{-7}$    &$3.63\times10^{-6}$    &$3.28\times10^{-6}$    &$7.18\times10^{-4}$\\
        $02$ &  $0.520$ & $12.000$  & $\pi/6$ & $2.00\times10^{-2}$ & $636.988$ &$2.10\times10^{-6}$    &$9.53\times10^{-7}$    &$1.48\times10^{-5}$    &$1.48\times10^{-5}$    &$1.78\times10^{-3}$\\
        $03$ &  $0.300$ & $14.400$  & $\pi/6$ & $2.00\times10^{-2}$ & $224.915$ &$9.39\times10^{-6}$    &$4.04\times10^{-6}$    &$3.92\times10^{-5}$    &$4.95\times10^{-5}$    &$5.03\times10^{-3}$\\
        $04$ &  $0.200$ & $16.800$  & $\pi/7$ & $2.72\times10^{-2}$ & $144.765$ &$4.60\times10^{-5}$    &$4.54\times10^{-6}$    &$7.31\times10^{-5}$    &$1.04\times10^{-4}$    &$7.37\times10^{-3}$\\
        $05$ &  $0.400$ & $16.800$  & $\pi/7$ & $2.72\times10^{-2}$ & $201.217$ &$1.33\times10^{-5}$    &$2.91\times10^{-6}$    &$6.17\times10^{-5}$    &$8.26\times10^{-5}$    &$5.29\times10^{-3}$\\
        $06$ &  $0.514$ & $27.243$  & $-\pi/12$ & $1.80\times10^{-2}$ & $30.518$ &$2.47\times10^{-4}$    &$1.96\times10^{-5}$    &$8.48\times10^{-4}$ &$1.39\times10^{-3}$    &$3.58\times10^{-2}$\\
        $07$ &  $0.500$ & $24.750$  & $\pi/4$ & $3.60\times10^{-2}$ & $59.061$ &$7.79\times10^{-5}$    &$3.33\times10^{-6}$    &$1.70\times10^{-4}$    &$3.44\times10^{-4}$    &$1.68\times10^{-2}$\\
        $08$ &  $0.600$ & $19.200$  & $\pi/5$ & $2.80\times10^{-2}$ &   $148.460$&$1.92\times10^{-5}$    &$1.91\times10^{-6}$    &$1.21\times10^{-4}$    &$1.92\times10^{-4}$    &$6.92\times10^{-3}$\\
        $09$ &  $0.700$ & $15.300$  & $\pi/6$ & $1.00\times10^{-2}$ & $140.487$ &$1.15\times10^{-5}$    &$4.43\times10^{-6}$    &$1.54\times10^{-4}$    &$1.80\times10^{-4}$    &$7.73\times10^{-3}$\\
        $10$ &  $0.800$ & $12.600$  & $\pi/8$ & $1.20\times10^{-2}$ & $355.391$ &$4.80\times10^{-6}$    &$3.42\times10^{-6}$    &$6.45\times10^{-5}$    &$6.18\times10^{-5}$    &$3.27\times10^{-3}$\\
        $11$ &  $0.100$ & $39.600$  & $-\pi/6$ & $7.00\times10^{-2}$ & $32.112$ &$5.38\times10^{-3}$    &$6.25\times10^{-5}$    &$3.66\times10^{-3}$    &$9.96\times10^{-3}$    &$5.57\times10^{-2}$\\
        $12$ &  $0.253$ & $35.093$  & $-\pi/3$ & $7.84\times10^{-2}$ & $39.303$&$7.72\times10^{-5}$    &$2.36\times10^{-7}$ &$4.13\times10^{-5}$    &$1.41\times10^{-4}$    &$2.63\times10^{-2}$\\
        $13$ &  $0.206$ & $30.159$  & $-\pi/4$ & $7.60\times10^{-2}$ & $45.575$ &$1.38\times10^{-4}$    &$1.94\times10^{-6}$    &$1.56\times10^{-4}$    &$3.69\times10^{-4}$    &$2.25\times10^{-2}$\\
        $14$ &  $0.368$ & $41.053$  & $-\pi/7$ & $8.00\times10^{-2}$ & $47.910$ &$1.49\times10^{-3}$    &$1.23\times10^{-5}$    &$3.00\times10^{-3}$    &$8.19\times10^{-3}$    &$3.61\times10^{-2}$\\
        $15$ &  $0.295$ & $47.924$  & $-\pi/9$ & $6.00\times10^{-2}$ & $24.535$ &$1.03\times10^{-2}$    &$1.18\times10^{-4}$    &$1.37\times10^{-2}$    &$3.34\times10^{-2}$    &$7.85\times10^{-2}$\\
        $16$ &  $0.425$ & $32.775$  & $-\pi/11$ & $3.00\times10^{-2}$ & $19.701$ &$6.08\times10^{-4}$    &$2.15\times10^{-5}$    &$1.04\times10^{-3}$    &$1.99\times10^{-3}$    &$5.27\times10^{-2}$\\
        $17$ &  $0.300$ & $27.300$  & $\pi/3$ & $3.20\times10^{-2}$ & $27.179$ &$9.05\times10^{-5}$    &$1.01\times10^{-6}$    &$7.17\times10^{-5}$    &$2.40\times10^{-4}$    &$3.78\times10^{-2}$\\
        $18$ &  $0.133$ & $44.200$  & $-2\pi/3$ & $6.80\times10^{-2}$ & $29.731$ &$9.80\times10^{-4}$    &$2.19\times10^{-6}$    &$4.83\times10^{-4}$    &$2.21\times10^{-3}$    &$3.43\times10^{-2}$\\
        $19$ &  $0.137$ & $50.039$  & $-\pi/3$ & $7.20\times10^{-2}$ & $24.736$ &$1.04\times10^{-3}$    &$2.53\times10^{-6}$    &$7.36\times10^{-5}$    &$2.42\times10^{-3}$    &$4.07\times10^{-2}$\\
        $20$ &  $0.477$ & $25.108$  & $-3\pi/5$ & $8.00\times10^{-2}$ & $101.345$ &$4.48\times10^{-5}$    &$2.05\times10^{-7}$    &$1.16\times10^{-5}$    &$3.58\times10^{-5}$    &$9.75\times10^{-3}$\\
        \hline
    \end{tabular}}
    \caption{Parameter setting and parameter estimation accuracy for the 20 X-MRIs at the GC}
    \label{tab:parameters}
\end{table*}

\begin{figure*}[!htbp]
\centering
\includegraphics[scale=0.45]{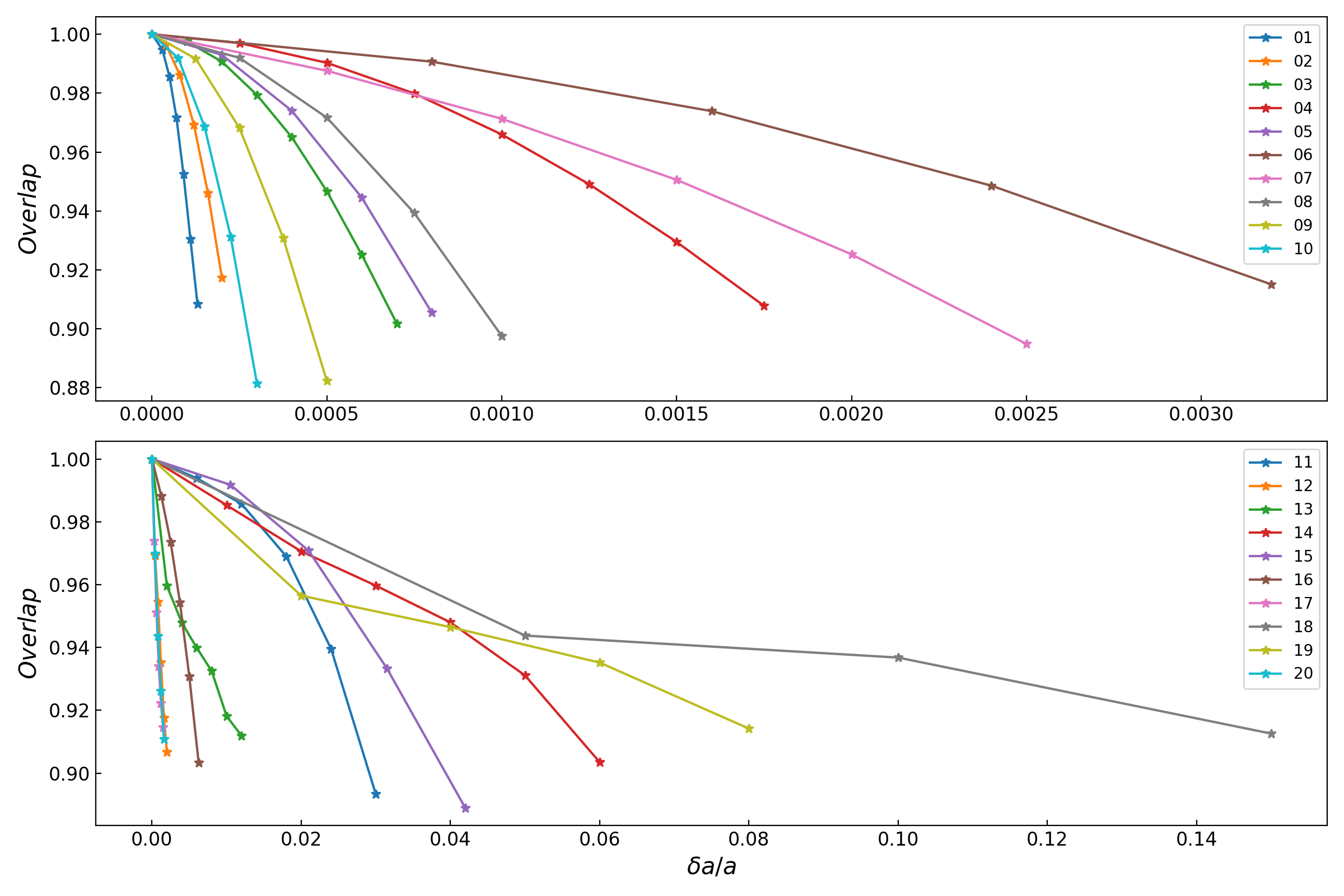}
\caption{Overlaps between the original waveforms and the waveforms changed with spin $a$. The other parameters ($M$, $\delta_{1}$, $\delta_{2}$, $e$, $p$, $\iota$) of systems listed in Table \ref{tab:parameters} remain unchanged. The top plane represents $a$-overlap curves from the top 10 systems (X-MRI 01 to X-MRI 10). The bottom plane represents $a$-overlap curves from the last 10 systems(X-MRI 11 to X-MRI 20).}
\label{fig:spin}
\end{figure*}

\begin{figure*}[!htbp]
\centering
\includegraphics[scale=0.45]{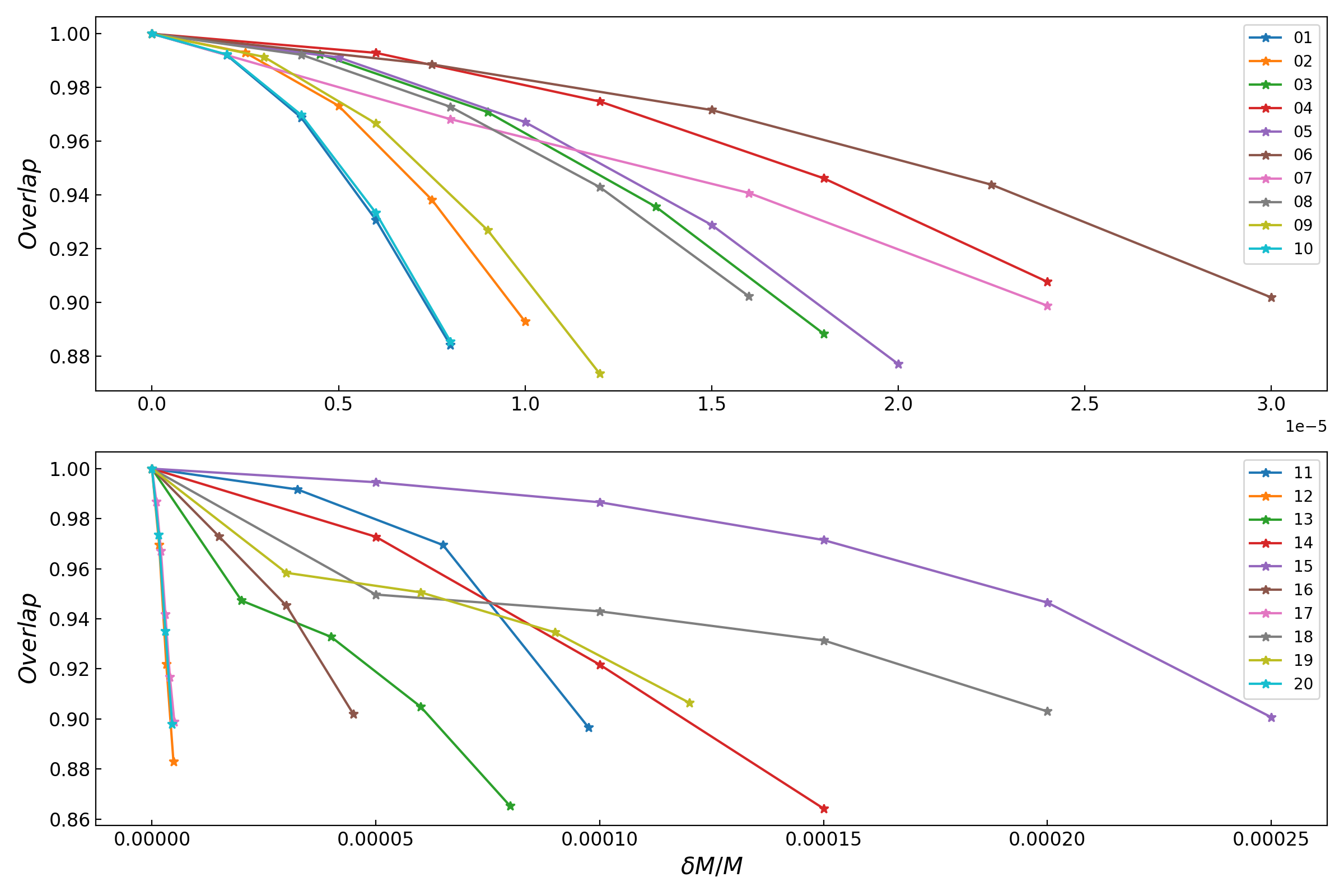}
\caption{Overlaps between the original waveforms and the waveforms changed with mass $M$. The other parameters ($a$, $\delta_{1}$, $\delta_{2}$, $e$, $p$, $\iota$) of systems listed in Table \ref{tab:parameters} remain unchanged. The top plane represents $M$-overlap curves from the top 10 systems(X-MRI 01 to X-MRI 10). The bottom plane represents $M$-overlap curves from the last 10 systems(X-MRI 11 to X-MRI 20).}
\label{fig:mass}
\end{figure*}

\begin{figure*}[!htbp]
\centering
\includegraphics[scale=0.45]{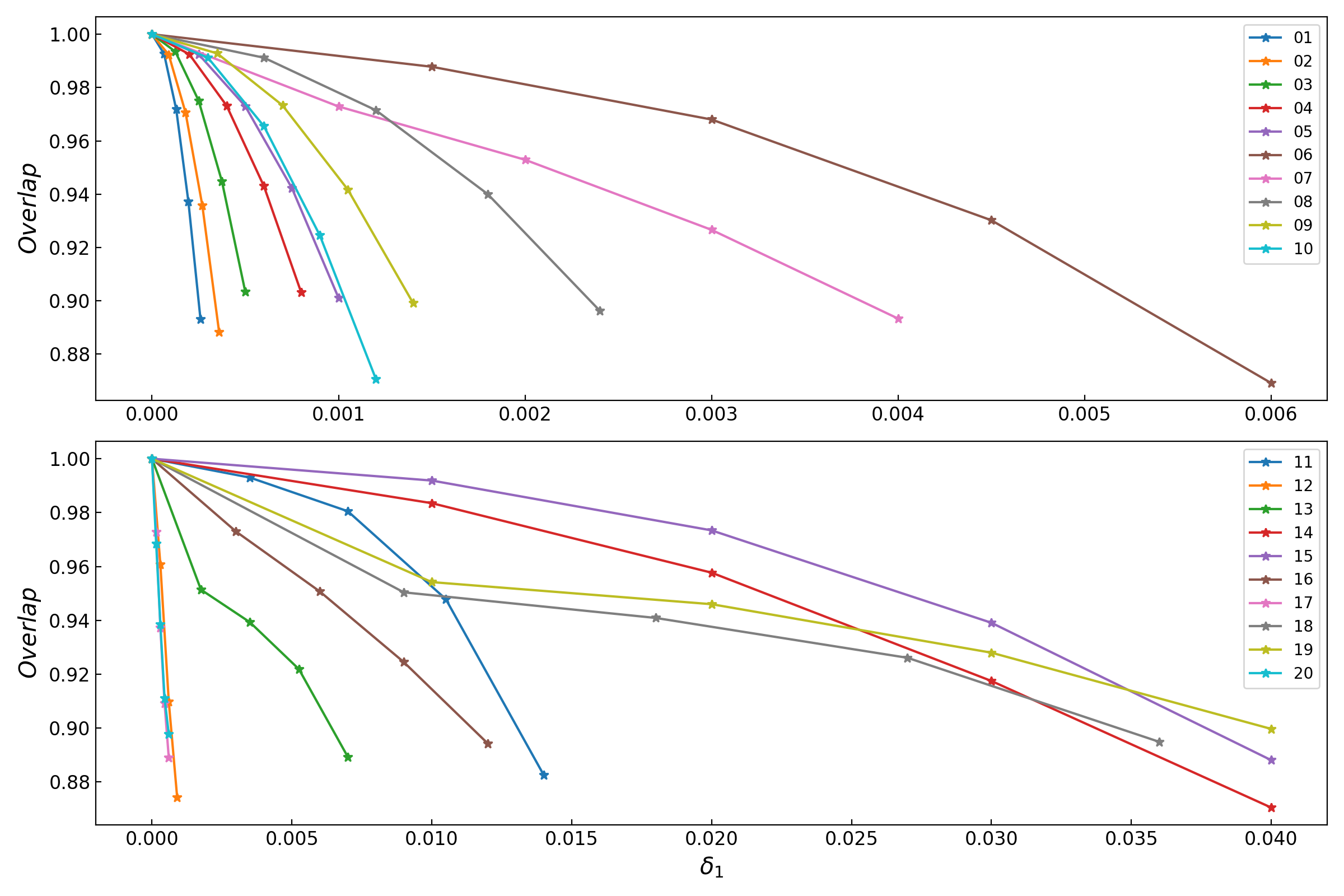}
\caption{Overlaps between the original waveforms and the waveforms changed with deformation parameter $\delta_{1}$. The other parameters ($M$, $a$, $\delta_{2}$, $e$, $p$, $\iota$) of systems listed in Table \ref{tab:parameters} remain unchanged. The top plane represents $\delta_{1}$-overlap curves from the top 10 systems(X-MRI 01 to X-MRI 10). The bottom plane represents $\delta_{1}$-overlap curves from the last 10 systems(X-MRI 11 to X-MRI 20).}
\label{fig:delta1}
\end{figure*}

\begin{figure*}[!htbp]
\centering
\includegraphics[scale=0.45]{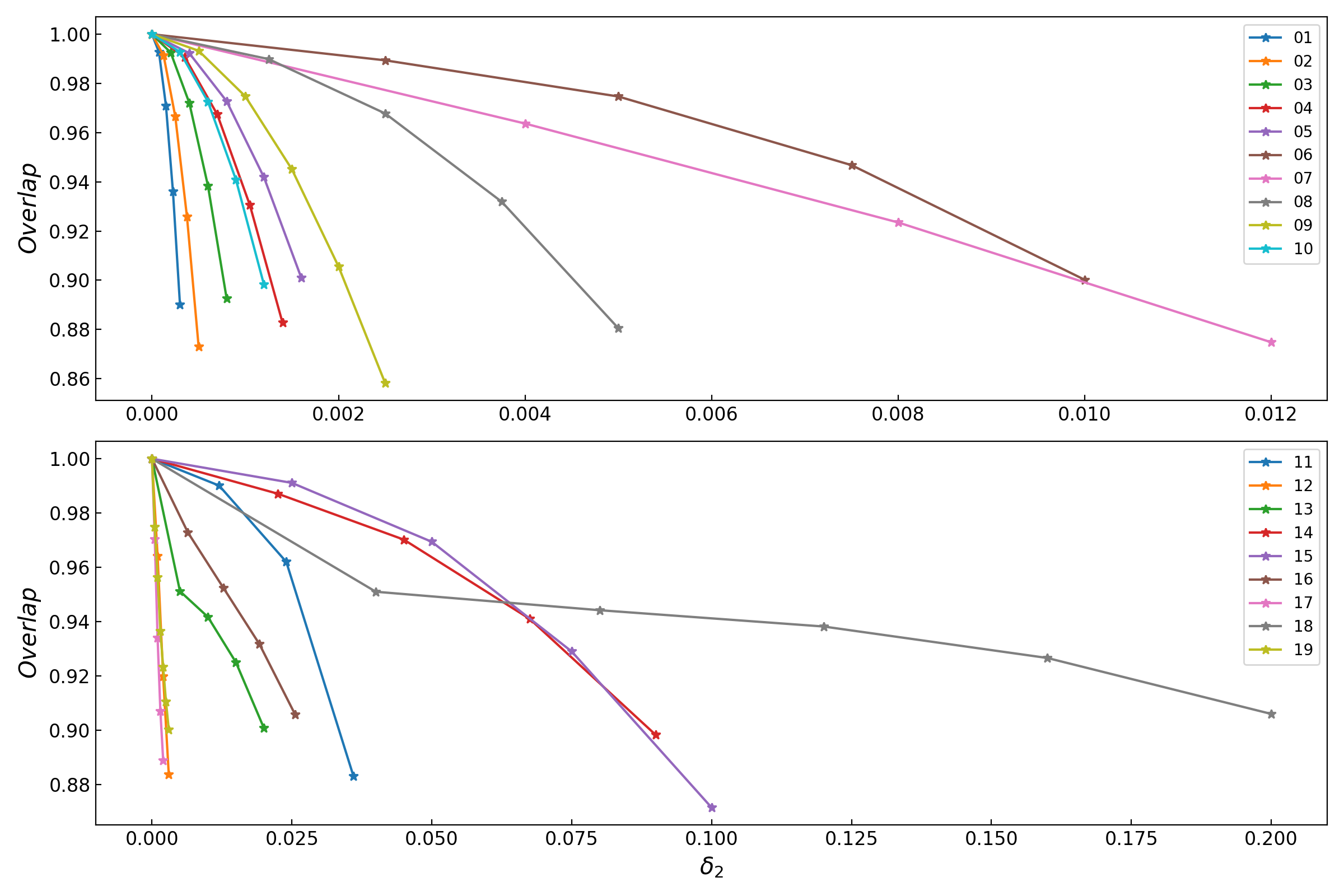}
\caption{Overlaps between the original waveforms and the waveforms changed with deformation parameter $\delta_{2}$. The other parameters ($M$, $a$, $\delta_{1}$, $e$, $p$, $\iota$) of systems listed in Table \ref{tab:parameters} remain unchanged. The top plane represents $\delta_{2}$-overlap curves from the top 10 systems(X-MRI 01 to X-MRI 10). The bottom plane represents $\delta_{2}$-overlap curves from the last 10 systems(X-MRI 11 to X-MRI 20).}
\label{fig:delta2}
\end{figure*}

Suppose the GW signal and corresponding GW template overlaps are above 0.97\cite{glampedakis2006mapping}. In that case, we would find neither the deviations from GR nor the unusual parameters of X-MRIs, which is called the confusion problem\cite{glampedakis2006mapping}. The confusion problem can prevent us from getting accurate parameter estimation of the X-MRIs. To make sure there is no confusion in our study, we calculate the overlaps between different gravitational waveforms of twenty X-MRIs with varying parameters $\lambda_i$ $(\lambda_i=a, M, \delta_{1}, \delta_{2}, e, p, \iota)$. Here $(a, M)$ are the parameters of the Sgr A*, $(\delta_{1}, \delta_{2})$ are the deformation parameters of the space-time from the Kerr solution, and $(e, p, \iota)$ are the parameters of orbit (eccentricity, semi-latus rectum, inclination).
  
Because $a, M, \delta_{1}$, and $\delta_{2}$ are the intrinsic parameters of Sgr A* and present the nature of MBH directly, we pay more attention to these four parameters. The Figs.~\ref{fig:spin}-\ref{fig:delta2} display the  overlaps between the original waveforms and the waveforms with varying parameters $a, M, \delta_{1}$ and $\delta_{2}$. As these figures show, the overlap tends to decrease while the increment of  $\lambda$ increases.

Taking the overlap value 0.97 as a criterion would give the constraints on $\lambda$.
Specifically, to get the constraints on $\delta\lambda_{i}$ by the GWs of X-MRI, we first keep the other parameters fixed and generate several waveforms with varying $\lambda_{i}$. Then we calculate the overlaps between the original waveform and the waveforms with varying $\lambda_{i}$. Finally, the corresponding value of $\lambda_{i}$ when overlap equals 0.97 can be regarded as the limit of $\lambda_{i}$.
From these figures, we observe the parameter constraint ability for different X-MRI varies.


\subsection{Evaluate the accuracy of parameter estimation for X-MRIs}
\label{B}
The SNRs of the X-MRI GW signals is high enough to apply the Fisher information matrix to estimate the accuracy of parameter estimation. We present the accuracy of parameter estimation for Sgr A* in this part using the Fisher information matrix. To better estimate the distance between Sgr A* and the solar system, we take account of the external parameter $R_{p}$ and constrain it by the gravitational waveforms of the X-MRIs in Table~\ref{tab:parameters}.

The Fisher information matrix $\Gamma$ for a GW signal $h$ parameterized by $\lambda$ is given by (See Ref\cite{cutler1994gravitational} for details)
\begin{equation}
\label{Gamma}
    \Gamma_{i,j}=<\frac{\partial h}{\partial\lambda_{i}}\vert\frac{\partial h}{\partial\lambda_{j}}>,
\end{equation}
where $\lambda_{i}=(a, M, \delta_{1}, \delta_{2}, e, p, \iota, R_{p})$ is one of the parameters of the X-MRI system. The parameter estimation uncertainty $\Delta\lambda$ due to Gaussian noise has the normal distribution $\mathcal{N}(0,\Gamma^{-1})$ in the case of high SNR, so the root-mean-square  uncertainty in the general case can be approximated as
\begin{equation}
\label{error}
    \Delta\lambda_{i}=\sqrt{(\Gamma^{-1})_{i,i}}. 
\end{equation}
For parameter estimation uncertainty $\Delta\lambda_{i},\Delta\lambda_{j}(i\neq j)$, the corresponding likelihood is \cite{cutler1994gravitational,babak2017science,han2019testing}.
\begin{equation}
\label{likelihood}
    \mathcal{L}(\lambda)\propto e^{-\frac{1}{2}\Gamma_{i,j}\Delta\lambda_{i}\Delta\lambda_{j}}. 
\end{equation}
For an X-MRI with eight parameters, we can get a Fisher matrix $(\Gamma_{i,j})_{8\times8}$ by applying the results of these parameters' preliminary constraints to equation (\ref{Gamma}). Element $\Gamma_{i,j}$ ($i\neq j$) in the Fisher matrix is the result of the combination of parameter $\lambda_{i}$ and parameter $\lambda_{j}$.  With the Fisher matrix, absolute uncertainty $\Delta\lambda_{i}$ of any parameter $\lambda_{i}$ can be estimated by calculating the equation (\ref{error}). Here we focus on the estimations of Sgr A*'s parameters $(a, M, \delta_{1}, \delta_{2}, R_{p})$. 

\begin{figure*}[!htbp]
\centering
\includegraphics[scale=0.48]{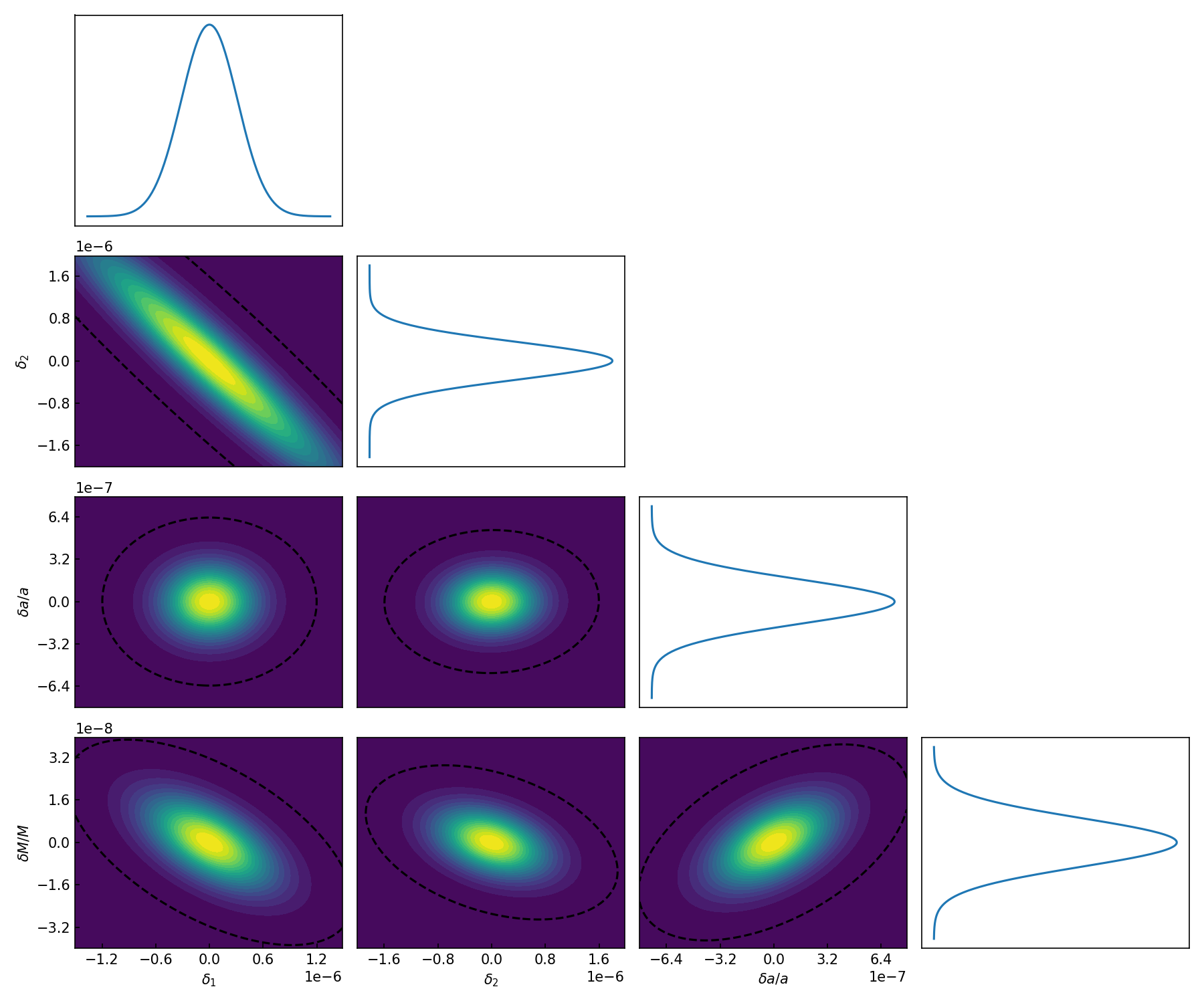}
\caption{Likelihoods of $(\delta_{1}, \delta M/M)$,  $(\delta_{1}, \delta a/a)$, $(\delta_{1}, \delta_{2})$, $(\delta_{2}, \delta M/M)$, $(\delta_{2}, \delta a/a)$, $(\delta a/a, \delta M/M)$ derived from the Fisher matrix of X-MRI 01. The black dashed eclipses show the $3\sigma$ confidence level. The upper and the four right-hand panels show the marginalized probability distribution for $\delta_{1}, \delta_{2}, \delta a/a$ and $\delta M/M$, respectively.}
\label{fig:EM01 01}
\end{figure*}

\begin{figure*}[!htbp]
\centering
\includegraphics[scale=0.48]{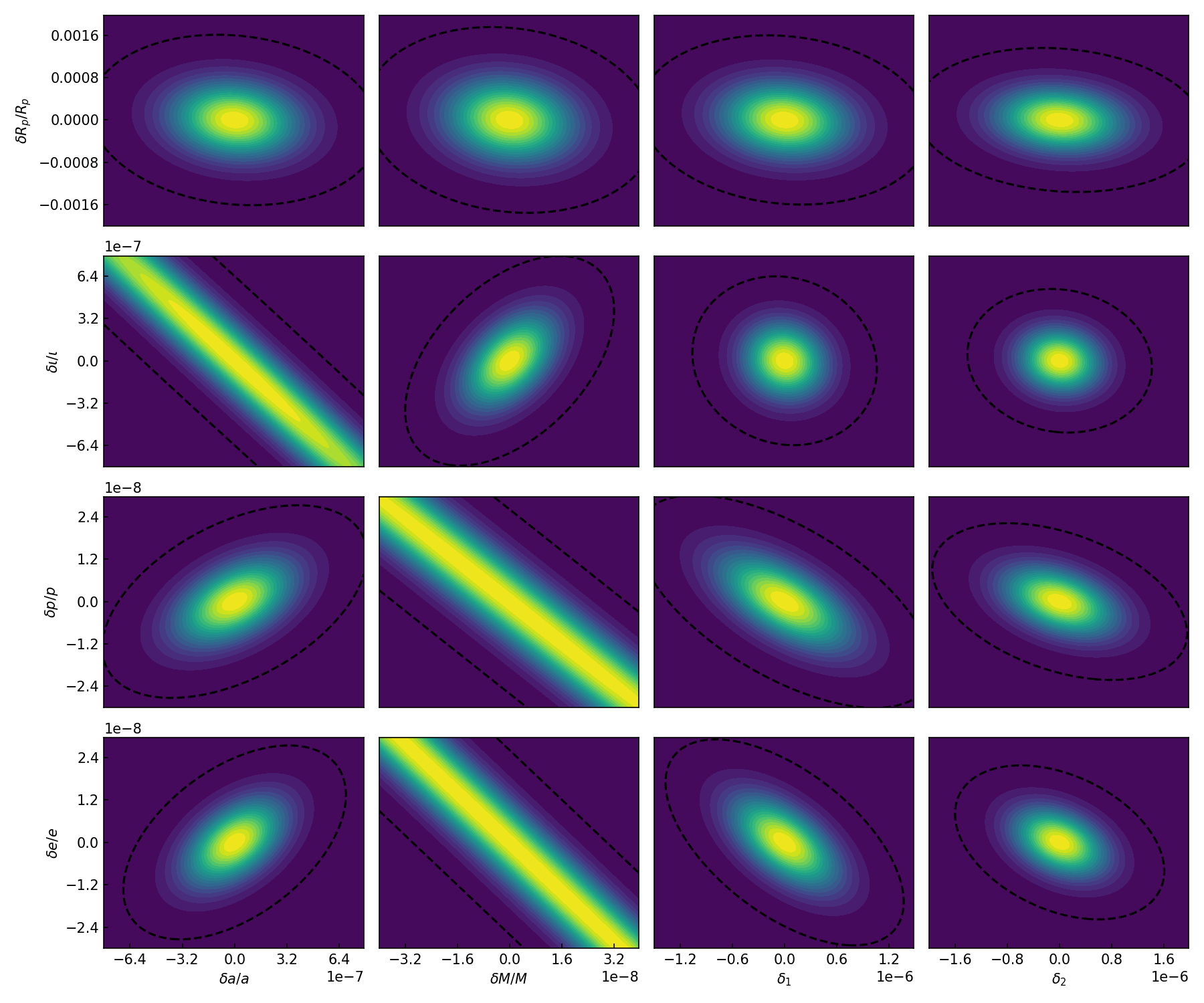}
\caption{Likelihoods of $(\delta a/a, \delta e/e)$,  $(\delta a/a, \delta p/p)$, $(\delta a/a, \delta\iota/\iota)$, $(\delta a/a, \delta R_{p}/R_{p})$, $(\delta M/M, \delta e/e)$,  $(\delta M/M, \delta p/p)$, $(\delta M/M, \delta\iota/\iota)$, $(\delta M/M, \delta R_{p}/R_{p})$, $(\delta_{1}, \delta e/e)$,  $(\delta_{1}, \delta p/p)$, $(\delta_{1}, \delta\iota/\iota)$, $(\delta_{1}, \delta R_{p}/R_{p})$, $(\delta_{2}, \delta e/e)$,  $(\delta_{2}, \delta p/p)$, $(\delta_{2}, \delta\iota/\iota)$, $(\delta_{2}, \delta R_{p}/R_{p})$ derived from the Fisher matrix of X-MRI 01. The black dashed eclipses show the $3\sigma$ confidence level.}
\label{fig:EM01 02}
\end{figure*}

\begin{figure*}[!htbp]
\centering
\includegraphics[scale=0.5]{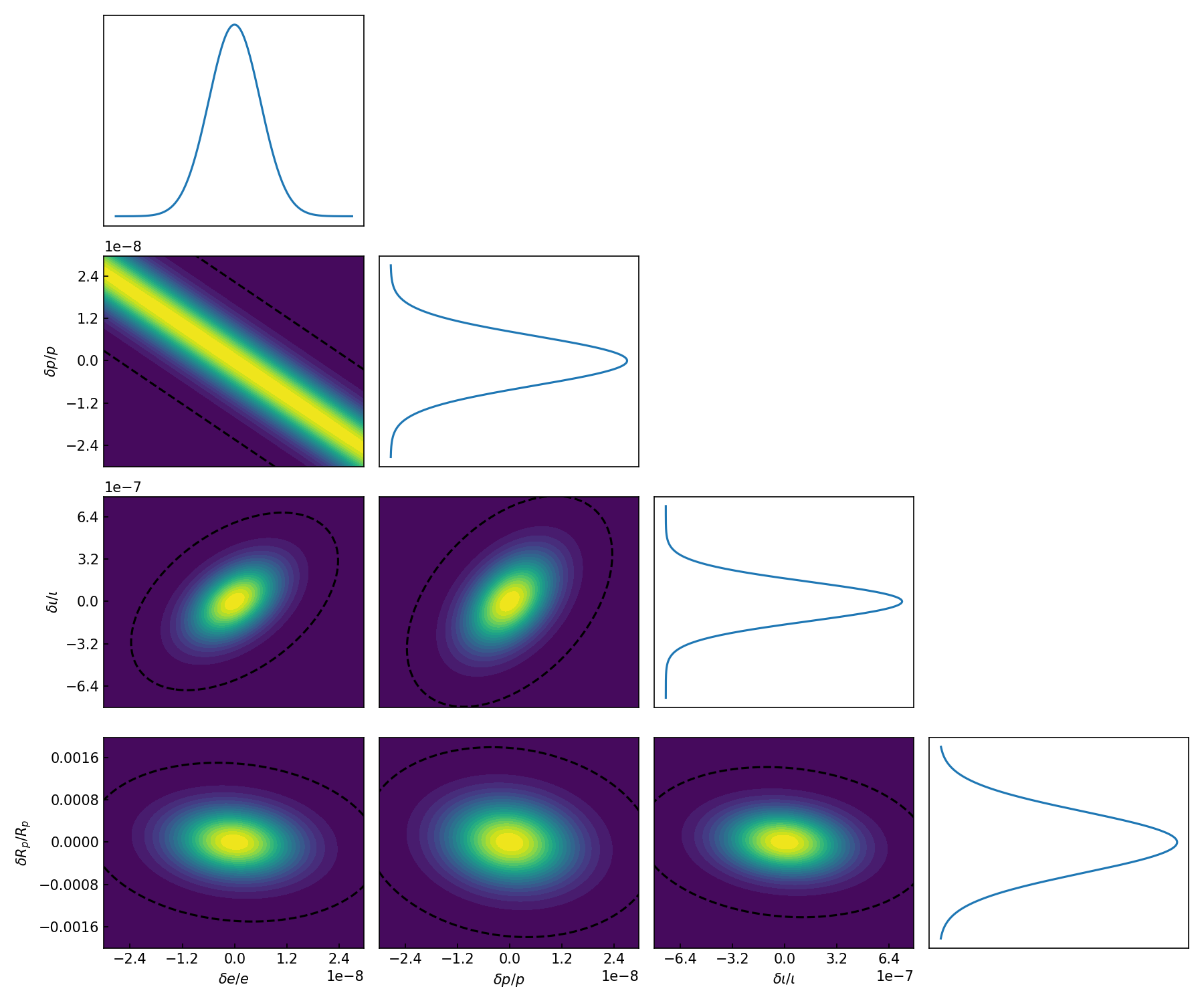}
\caption{Likelihoods of $(\delta e/e, \delta R_{p}/R_{p})$,  $(\delta e/e, \delta\iota/\iota)$, $(\delta e/e, \delta p/p)$, $(\delta p/p, \delta R_{p}/R_{p})$, $(\delta p/p, \delta\iota/\iota)$, $(\delta\iota/\iota, \delta R_{p}/R_{p})$ derived from the Fisher matrix of X-MRI 01. The black dashed eclipses show the $3\sigma$ confidence level. The upper and the four right-hand panels show the marginalized probability distribution for $\delta e/e, \delta p/p, \delta\iota/\iota$ and $\delta R_{p}/R_{p}$, respectively.}
\label{fig:EM01 03}
\end{figure*}

By using the Fisher matrix, the parameter estimation accuracy of $(a, M, \delta_{1}, \delta_{2}, R_{p})$ for the twenty X-MRI signals is shown in Table \ref{tab:parameters}. Different X-MRI systems have different abilities to estimate the uncertainty accuracy of the same parameter. For the spin of Sgr A*, the relative uncertainty $\Delta a/a$ estimated by X-MRI 01, X-MRI 02, X-MRI 03, and X-MRI 10 reach a very high precision $\sim10^{-6}$. While $\Delta a/a$ estimated by X-MRI 15 is only $\sim10^{-2}$. For the mass of Sgr A*, its relative uncertainty $\Delta M/M$ estimated by X-MRI 01, X-MRI 02, X-MRI 12, and X-MRI 20 reach $\sim10^{-7}$, and $\Delta M/M$ estimated by X-MRI 15 is $\sim10^{-4}$. For the space-time deformation around Sgr A*, $\Delta\delta_{1}$ and $\Delta\delta_{2}$ estimated by X-MRI 01 reach $\sim10^{-6}$, while the relative uncertainty of these deformation parameters estimated by X-MRI 15 is only $\sim10^{-2}$. For the distance $R_{p}$, its relative uncertainty $\Delta R_{p}/R_{p}$ estimated by X-MRI 01 reaches $\sim10^{-4}$,  while the accuracy of $\Delta R_{p}/R_{p}$ estimated by X-MRI 06, X-MRI 07,  X-MRI 11, and X-MRI 19 is only $\sim10^{-2}$. From the above analysis, we find that X-MRI 01 has stringent constraints for the five parameters $(a, M, \delta_{1}, \delta_{2}, R_{p})$. Therefore, we take X-MRI 01 as an example to present its likelihoods calculated by Eqs. \ref{Gamma}-\ref{likelihood}. As shown in Figs.~\ref{fig:EM01 01}-\ref{fig:EM01 03}, it is obvious that the parameter estimation for X-MRI 01 may be affected by any other parameter. Thus, it is reasonable to consider the parameters of one X-MRI signal to estimate any parameter.

\begin{figure*}[!htbp]
\centering
\includegraphics[scale=0.4]{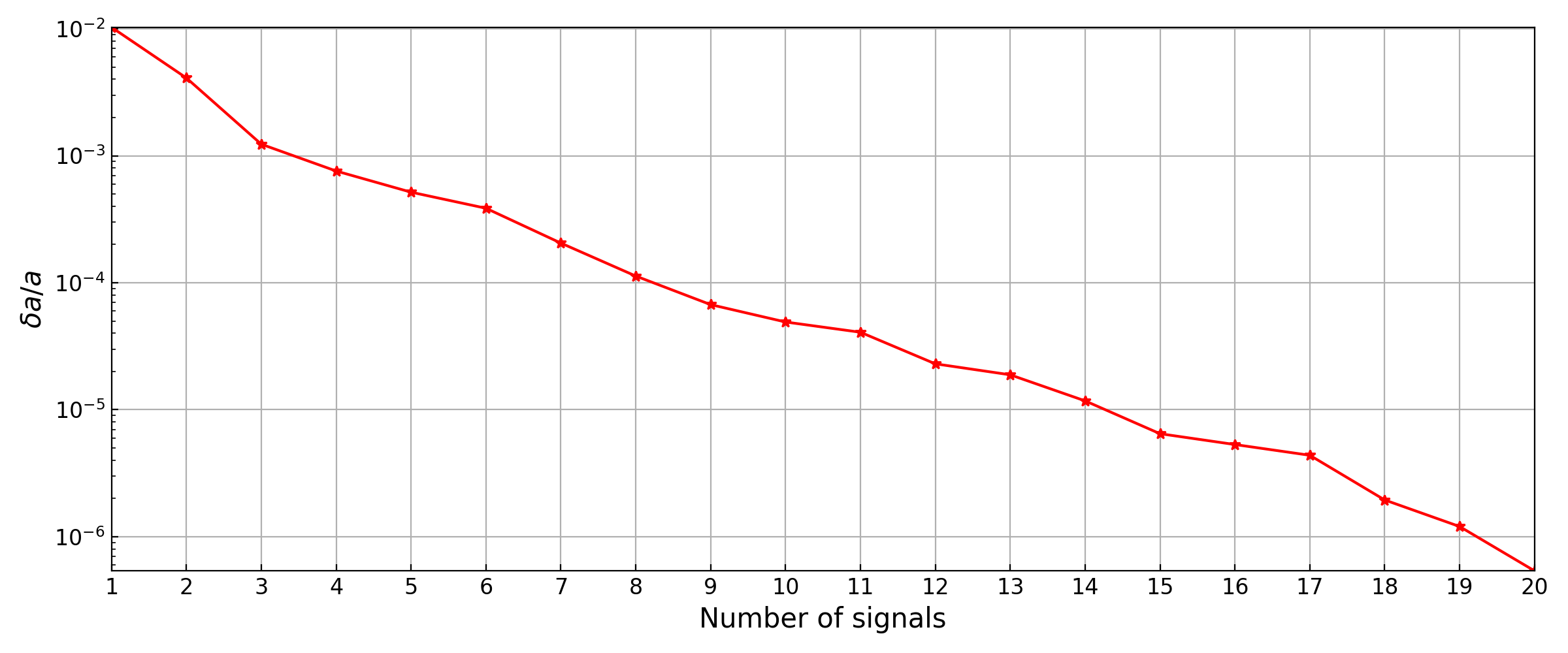}
\includegraphics[scale=0.4]{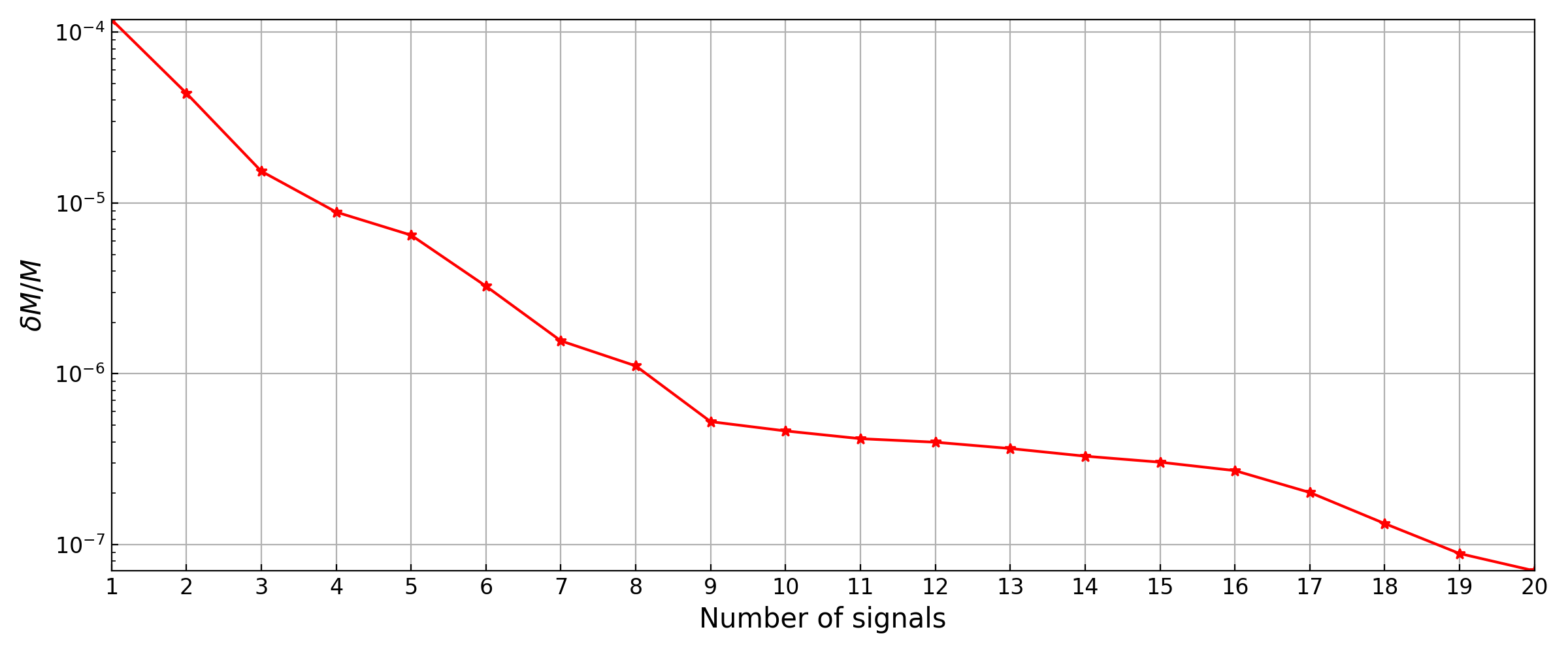}
\includegraphics[scale=0.4]{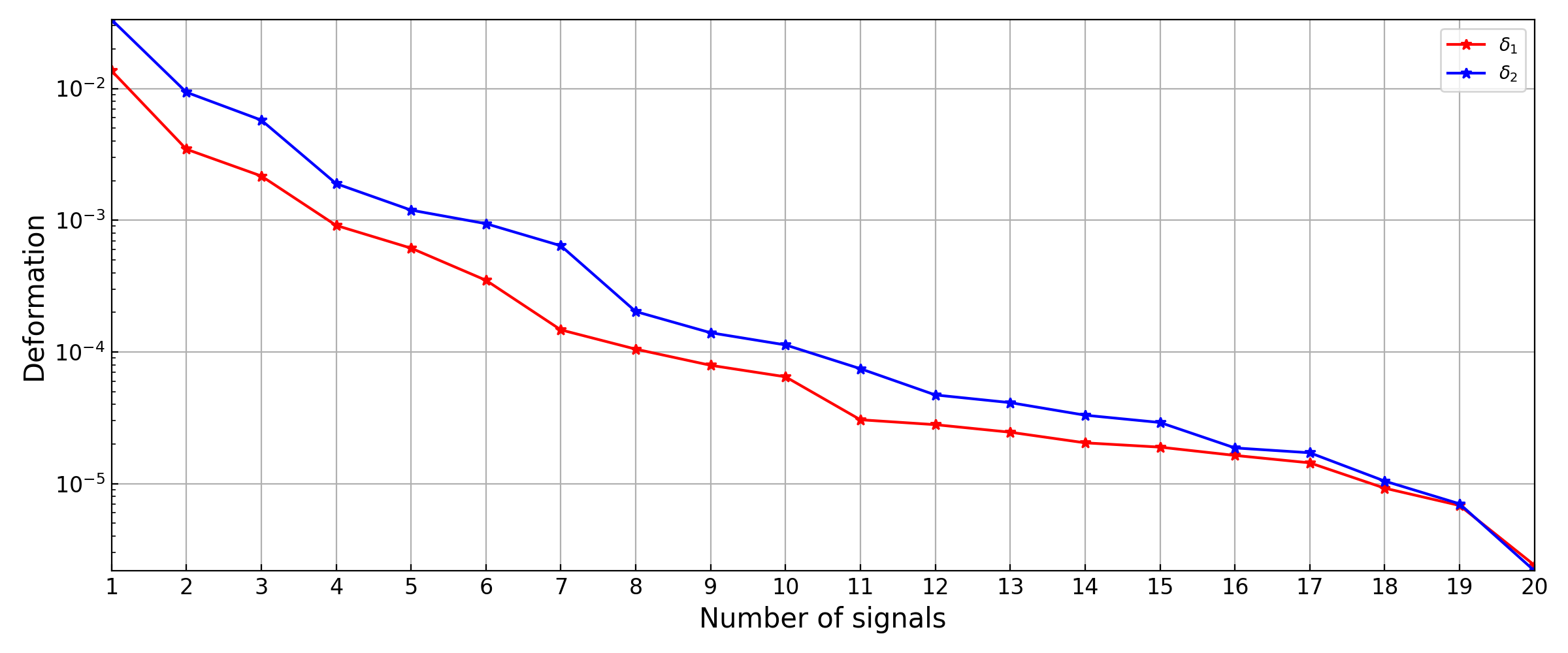}
\includegraphics[scale=0.4]{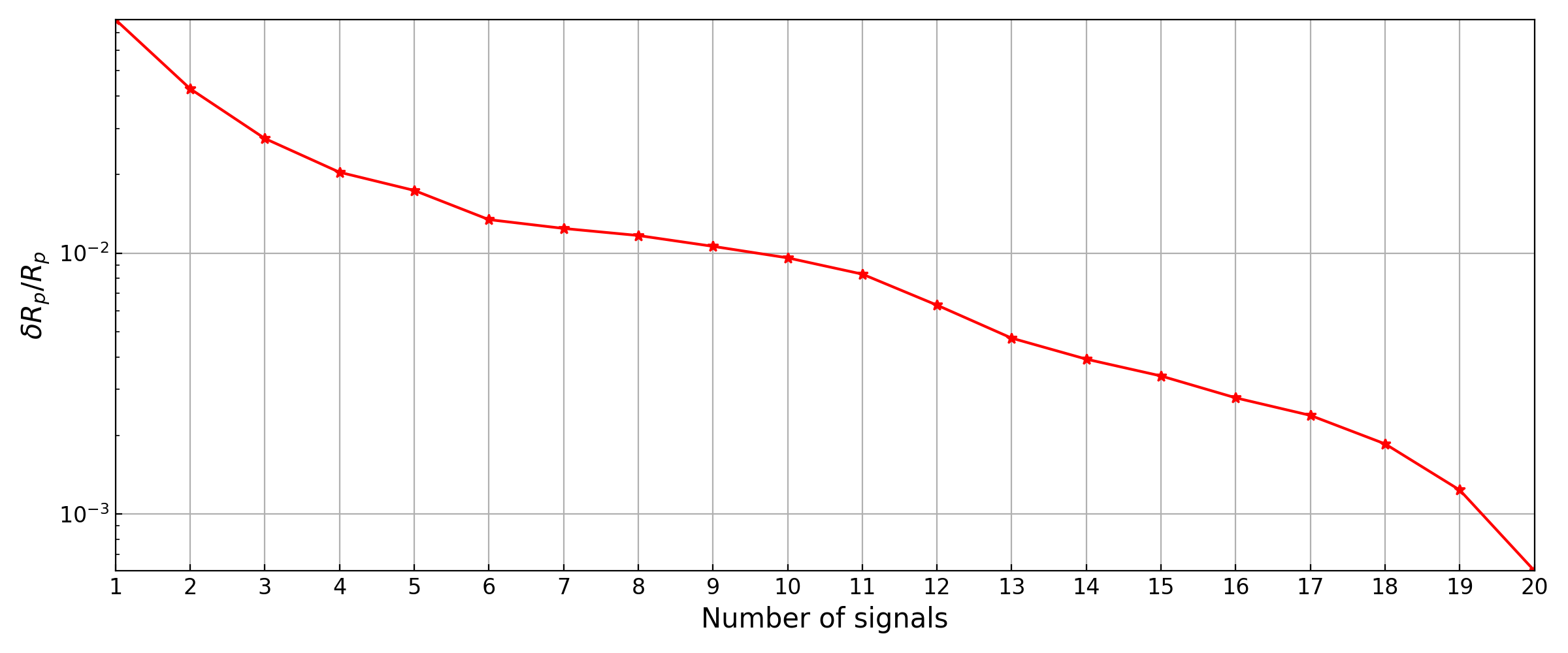}
\caption{The relation between the parameter estimation accuracy and the X-MRI signals number.
The parameter in the first plane is $\delta a/a$, in the second plane is $\delta M/M$, in the third plane are $\delta_{1}$(red) and $\delta_{2}$(blue), in the fourth plane is $\delta R_{p}/R_{p}$.}
\label{fig:A CUM}
\end{figure*}

We further study the influence of the combination of GW signals on the parameter estimation accuracy. Here we take parameter $\alpha$ as an example to present the data processing. Firstly, we assume that there are $n$ X-MRI systems at the GC. Then, we calculate the Fisher matrices of all these signals to determine the diagonal element $\Gamma_{\alpha,\alpha}$. Sort the value of $\Gamma_{\alpha,\alpha}$ by the order of size, and the corresponding matrix will be $\Gamma_{\alpha 1}, \Gamma_{\alpha 2}, ..., \Gamma_{\alpha n}$. Then we add these matrices to get the matrix $\Gamma_{\alpha}$,
\begin{equation}
\label{Gamma_alpha}
    \Gamma_{\alpha}=\Gamma_{\alpha 1}+\Gamma_{\alpha 2}+...+\Gamma_{\alpha n},
\end{equation}
with $\Gamma_{\alpha}$, we get the estimation of absolute uncertainty from the equation
\begin{equation}
\label{estimation_alpha}
    \Delta\alpha=\sqrt{(\Gamma_{\alpha})_{\alpha,\alpha}^{-1}}. 
\end{equation}
We repeat the steps of the estimation for $\Delta\alpha$, and calculate the absolute uncertainty of $a, M, \delta_{1}, \delta_{2}, R_{p}$. Then we will get the relative uncertainty. The results are shown in Figs.~\ref{fig:A CUM}. The accuracy gets better as the number of X-MRI increases. With all twenty X-MRI systems in Table \ref{tab:parameters}, the estimation accuracy for these parameters all reach higher precision. $\Delta a/a$ reaches the accuracy $\sim10^{-7}$. $\Delta M/M$ reaches the accuracy $\sim10^{-8}$. $\Delta\delta_{1}$ reaches the accuracy $\sim10^{-6}$. $\Delta\delta_{2}$ reaches the accuracy $\sim10^{-6}$. $\Delta R_{p}/R_{p}$ reaches the accuracy $\sim10^{-4}$. The observation number of X-MRI systems does make sense for parameter estimation. Finally, we must emphasize that the parameter estimation results predicted by the Fisher information matrix here only stand for the ideal situation, in the actual parameter estimation practice, because of all kinds of noise, the results would not be that kind of good.

\begin{table}[!htbp]
    \centering
    \begin{tabular}{ccccc}
    \toprule
        $\Delta a/a$&$\Delta M/M$&$\Delta\delta_{1}$&$\Delta\delta_{2}$&$\Delta R_{p}/R_{p}$\\
        \hline
        $5.38\times10^{-7}$     &$7.02\times10^{-8}$    &$2.40\times10^{-6}$    &$2.18\times10^{-6}$    &$6.05\times10^{-4}$\\
        
        \hline
    \end{tabular}
    \caption{Results of parameter estimation accuracy of all the 20 X-MRI systems.}
    \label{tab:results_in}
\end{table}

\section{Conclusions and Outlook}
\label{sec:conclusion}

Sgr A* is the closest MBH for the Solar system. It is therefore an ideal laboratory to study the properties of black holes and to test alternative theories of gravity. To investigate the structure of Sgr A*, we simulate the GW signals for twenty X-MRI systems using the KRZ metric and the kludge waveform. We then apply the Fisher information matrix method to these GW signals. With a single GW X-MRI event detected, we were able to obtain a relatively accurate estimate of spin $a$, mass $M$, and deviation parameters $\delta_{1}, \delta_{2}$. More X-MRI observations would improve the measurement of the above parameters.

In practice, galactic binaries(GBs) and EMRIs are also promising sources of space-borne GW detectors like LISA\cite{amaro2017laser}. GBs, comprise primarily white dwarfs but also neutron stars and stellar-origin black holes, emit continuous and nearly monochromatic GW signals. X-MRIs can be also regarded as monochromatic sources for space-borne detectors, while the signals of X-MRIs could reach high SNRs, making X-MRIs feasible to be distinguished from weaker sources such as GBs\cite{amaro2019extremely}. On the contrary, EMRIs, which evolve relatively rapidly, are polychromatic sources\cite{amaro2019extremely}. Therefore, EMRIs and X-MRIs could be complementary in studying the space-time of MBH.

\authorcontributions{conceptualization, Shu-Cheng Yang; methodology, Huijiao Luo, Yuan-Hao Zhang, Chen Zhang and Shu-Cheng Yang; software, Huijiao Luo, Yuan-Hao Zhang and Shu-Cheng Yang; validation, Shu-Cheng Yang; formal analysis, Huijiao Luo, Yuan-Hao Zhang and Shu-Cheng Yang; investigation, Huijiao Luo, Yuan-Hao Zhang Chen Zhang, and Shu-Cheng Yang; resources, Shu-Cheng Yang; data curation, Huijiao Luo, Yuan-Hao Zhang and Shu-Cheng Yang; writing–original draft preparation, Huijiao Luo; writing–review and editing, Shu-Cheng Yang and Yuan-Hao Zhang; visualization, Huijiao Luo ; supervision, Shu-Cheng Yang; project administration, Shu-Cheng Yang; funding acquisition, Shu-Cheng Yang. All authors have read and agreed to the published version of the manuscript.}

\funding{This work is supported by The National Key R\&D Program
of China (Grant No. 2021YFC2203002), NSFC (National Natural Science Foundation of China) No. 11773059 and No. 12173071.}

\acknowledgments{We thank Dr. Ahmadjon Abdujabbarov and Dr. Imene Belahcene for their valuable advice on this work.}

\conflictsofinterest{The authors declare no conflict of interest.} 




\abbreviations{Abbreviations}{
The following abbreviations are used in this manuscript:\\

\noindent 
\begin{tabular}{@{}ll}
FF & fitting factor\\
GC & Galactic Center\\
GW & gravitational wave\\
GR & general relativity\\
LIGO & Laser Interferometer Gravitation Wave Observatory\\
LISA & Laser Interferometer Space Antenna\\
MBH & massive black hole\\
SNR & signal-to-noise ratio\\
X-MRI & extremely large mass-ratio inspiral\\

\end{tabular}
}




\begin{adjustwidth}{-\extralength}{0cm}

\reftitle{References}

\end{adjustwidth}
\end{document}